\documentclass[pra,twocolumn,amsmath,amssymb,superscriptaddress,eqsecnun]{revtex4-2}
\usepackage{graphicx,amsmath,relsize,epstopdf,color,mathtools,bm,newtxtext,newtxmath,braket,rotating,float}
\usepackage[hyphenbreaks]{breakurl}
\usepackage[colorlinks=true,linkcolor=blue,citecolor=blue,urlcolor =blue]{hyperref}
\usepackage[normalem]{ulem}
\usepackage[table,xcdraw]{xcolor}
\newcommand{\RE}{\mathop{\mathrm{Re}} \nolimits}

\newcommand{\rect}{\mathop{\mathrm{rect}} \nolimits}
\newcommand{\erfi}{\mathop{\mathrm{erfi}} \nolimits}

\begin{document}

\title{A quantum trajectory analysis of singular wave functions}

\author{Angel S. Sanz}
\affiliation{Departamento de \'{O}ptica, Facultad de F\'{\i}sica, Universidad Complutense, 28040~Madrid, Spain}

\author{Luis~L. S\'{a}nchez-Soto}
\affiliation{Departamento de \'{O}ptica, Facultad de F\'{\i}sica, Universidad Complutense, 
28040~Madrid, Spain}
\affiliation{Max-Planck-Institut f\"{u}r die Physik des Lichts, 91058 Erlangen, Germany}

\author{Andrea Aiello}
\affiliation{Max-Planck-Institut f\"{u}r die Physik des Lichts, 91058 Erlangen, Germany}

\date{\today}

\begin{abstract}
The Schr\"{o}dinger equation admits smooth and finite solutions that spontaneously evolve into a singularity, even for a free particle. This blowup is generally ascribed to the intrinsic dispersive character of the associated time evolution. We resort to the notion of quantum trajectories to reinterpret this singular behavior. We show that the blowup can be directly related to local phase variations, which generate an underlying velocity field responsible for driving the quantum flux toward the singular region. 
\end{abstract}

\maketitle


\section{Introduction}
The Schr\"{o}dinger equation is, perhaps, the prototype of a dispersive equation; that is, if no boundary conditions are imposed, its wave solutions spread out in space as they evolve in time~\cite{Tao:2006aa}.  A frequent way to quantify this dispersion is by the so-called dispersive estimates, a topic with a long history~\cite{Schlag:2007aa,Mandel:2020aa,Dietze:2021aa} and whose main goal is to establish tight bounds on the decay of the solutions.  

Recently, it has been pointed out that the Schr\"{o}dinger equation, even for a free particle, presents dispersive singularities~\cite{Peres:2002aa,Bona:2010aa}: an initial square-integrable profile $\psi(x,0)$ could result in a solution $\psi(x, t)$ that blows up in a finite time.  In the remainder such profiles will be termed as \emph{singular wave packets}. While this singular behavior (sometimes denoted as self-focusing or wave collapse) is well understood in presence of nonlinearities~\cite{Sulem:1999aa,Fibich:2015aa,Karjanto:2020aa}, it is, at first sight, surprising in a pure linear evolution. 

From a mathematical viewpoint, this dispersive blowup can be related to the fact that the linear Schr\"{o}dinger equation is ill-posed in the space $L^{\infty}$: the free propagator is not a Fourier multiplier in $L^{\infty}$~\cite{Hormander:1960aa}. In physical terms, dispersive blowup is a focusing phenomenon due to both the unbounded domain of the problem and the propensity of the dispersion relation to propagating energy at different speeds. Interestingly, the same singular behavior has been described in for paraxial beams~\cite{Aiello:2016aa,Aiello:2020aa,Porras:2021aa}, which is consequent with the complete equivalence between the time-dependent Schr\"{o}dinger equation and the paraxial wave equation~\cite{Nienhuis:2017aa}.

In this paper, we address the physical interpretation of these singularities from the perspective of quantum trajectories. In this picture, quantum formalism is reinterpreted as describing particles following definite trajectories, each with a precisely defined position at each instant in time. However, in this approach, called Bohmian mechanics~\cite{Bohm:1952aa,Bohm:1952ab,Bohm:1993aa}, the trajectories of the particles are quite different from those of classical particles because they are guided by the wave function~\cite{Englert:1993aa,Sanz:2012aa,Mahler:2016aa,Sanz:2019aa}.  Our analysis shows that the blowup can be directly related to local phase variations, which generate an underlying velocity field (the phase gradient) responsible for driving the quantum flux toward the singular region. To shed light on this point, we compare the blowup with the focusing of a Gaussian and a rectangular wave packet: this demonstrates that imploding solutions are distinguished by an initial phase factor. 

Furthermore, for Gaussian wave packets, which can be nicely analyzed in closed form,  it is also observed that there are two types of solutions with very different properties, despite their initial density distributions being identical. One of such solutions leads to a classical type of propagation because the phase factor plays a minor role (or even no role at all). In contradistinction, the other type of solution is characterized by wide initial wave functions with an intrinsic highly oscillatory behavior.  This emphasizes the prominent role of the phase as an active agent in the subsequent dynamics.

This article is organized as follows. In Sec.~\ref{sec:blowup} we briefly discuss the spontaneous generation of a singularity in the Schr\"{o}dinger equation and introduce the basic elements needed to define a quantum trajectory. In terms of this notion, we analyze the singularity and put forward the fundamental role played by the quantum phase to understand that phenomenon. In Sec.~\ref{sec:wavepack} we examine the behavior of a Gaussian and a rectangular packet and compare with the previous singular wave.  Finally, Sec.~\ref{sec:conc} summarizes our conclusions.

\section{Dispersive blowup in the Schr\"{o}dinger equation}
\label{sec:blowup}

\subsection{Spontaneous generation of a singularity}

We first set the stage for our discussion. We will be considering the simplest case of the Schr\"{o}dinger equation for a free particle of mass $m$ in one dimension
\begin{equation}
\label{eq:Schr}
i \hbar \frac{\partial \psi (x,t)}{\partial t} = - \frac{\hbar^2}{2m} \frac{\partial^{2} \psi (x,t)}{\partial x^{2}} ,
\end{equation}
with the initial Cauchy problem $\psi (x,0 ) \in L^{2} (\mathbb{R})$. The unique solution of \eqref{eq:Schr} can be written in terms of the free-space propagator as~\cite{Merzbacher:1998aa}
\begin{equation}
\label{eq:psit}
\psi (x,t) = \sqrt{\frac{m}{2 \pi i \hbar t}} 
\int_{\mathbb{R}}  \exp \left [  \frac{i m}{2 \hbar t}(x- x^{\prime})^{2}\right ] \; 
\psi (x^{\prime},0) \, dx^{\prime}
\, ,
\end{equation}
where the integral has to be understood in the improper Riemann sense. In this way, the Schr\"{o}dinger equation appears as an integral equation, rather than a differential one, with the advantage of being valid even if the wave function is not a differentiable function.

Slightly generalizing results from Peres~\cite{Peres:2002aa}, we choose the initial data to be
\begin{equation}
\label{eq:psi0}
\psi (x,0) = \displaystyle \frac{1}{\sqrt{\mathcal{N}_{\nu}}} \frac{\exp \left (- \frac{i m}{2 \hbar \tau} x^{2}  \right )}{\left (1 + \frac{x^{2}}{\sigma^{2}} \right )^{\nu}} \, ,
\end{equation}
where $\mathcal{N}_{\nu}$ is a normalization constant, and  $\tau$ and $\sigma$  are real numbers fixing the time scale and the width of the distribution, respectively. One can check that for $\nu > 1/4$, this function is in the space  $L^{2} (\mathbb{R})$, and so it is a physically admissible solution. When this holds true, the normalization constant is finite and equal to $\mathcal{N}_{\nu} = \sqrt{\pi} \sigma \Gamma (2\nu - 1/2)/\Gamma ( 2\nu)$.

\begin{figure}[!t]
  \centering
  \includegraphics[width=.95\columnwidth]{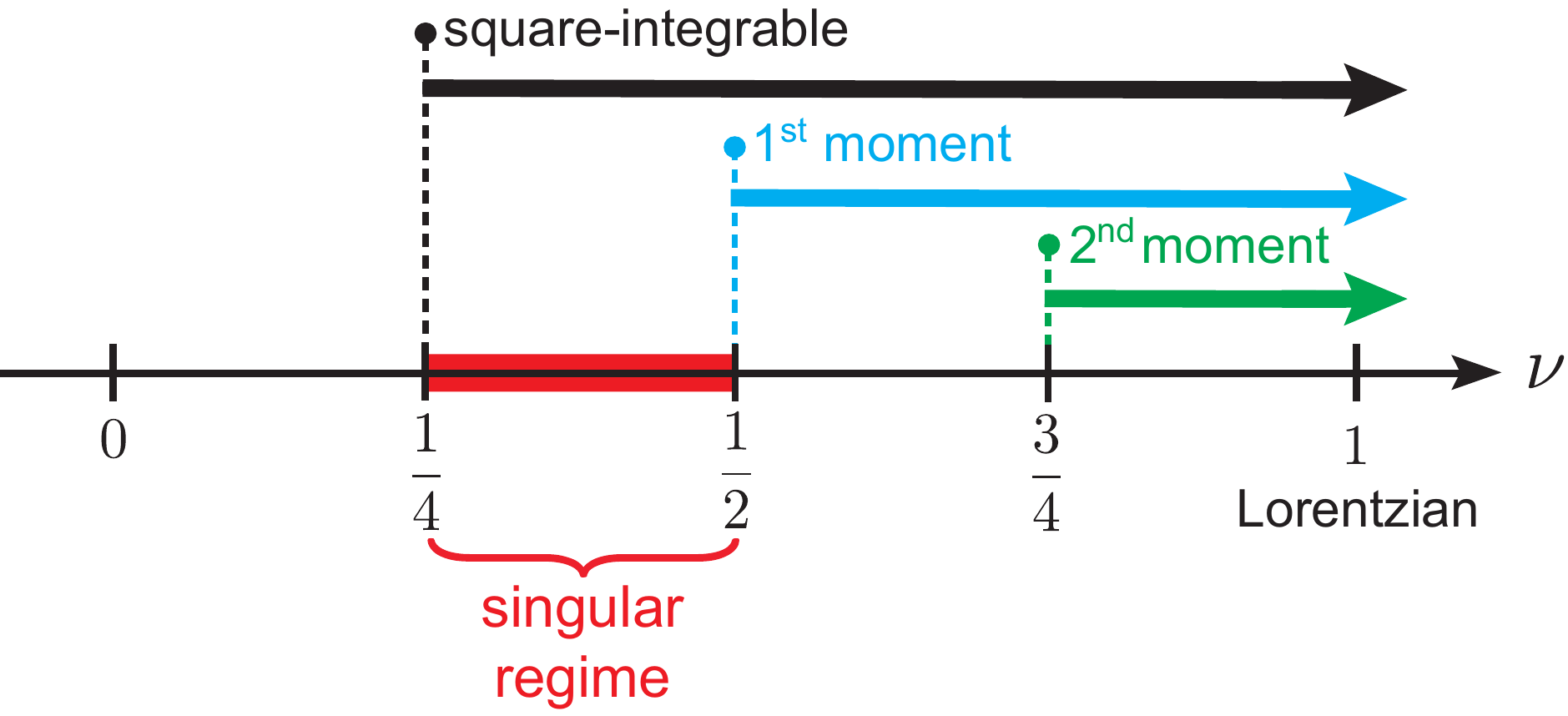}
    \caption{\label{fig1}
  For $\nu > 1/4$ the wave function $\eqref{eq:psi0}$ is square integrable. The red band indicates the range $1/4 < \nu < 1/2$ where the corresponding $\psi (x, t)$ exhibits a singularity at time $t=\tau$. For $\nu > 1/2$, the corresponding $\psi (x, t)$ is finite everywhere and the first moment $\langle x \rangle$ of the associated probability density $| \psi (x, t)|^{2}$ exists and it is equal to 0. Finally, the second moment $\langle x^{2} \rangle$ is finite for $\nu > 3/4$. The case $\nu=1$ corresponds to the Lorentzian function.}
\end{figure}

For $t \neq \tau$, we can apply the Riemann-Lebesgue lemma~\cite{Iorio:2001aa} to show that the resulting $\psi (x,t)$ is continuous in $x$ and $t$ and tends to zero as $|x | \rightarrow \infty$ (although not necessarily uniformly with respect to $t$). However, at $t=\tau$ a discontinuity occurs: at this time the wave function reads
\begin{equation}
\psi (x, \tau) = \sqrt{\frac{m}{2 \pi i \hbar \mathcal{N}_{\nu}}} e^{ \frac{i m}{2 \hbar \tau} x^{2}}   \int_{\mathbb{R}} \frac{e^{- i \frac{m}{\hbar \tau} x x^{\prime}}}{\left (1 + x^{\prime 2}/\sigma^{2} \right )^{\nu}} dx^{\prime} \, .
\end{equation}
This integral is the Fourier transform of a Bessel potential~\cite{Aronszajn:1961aa} and can thus be expressed as 
\begin{align}
\psi (x, \tau) = \sqrt{\frac{m \sigma^{2}}{i \hbar \mathcal{N}_{\nu}}} 
\frac{e^{\frac{i m}{2 \hbar \tau } x^{2}}}{2^{\nu -1} \Gamma (\nu)} \left ( \frac{m \sigma}{\hbar \tau} | x | \right )^{\nu - \case{1}{2}}    
K_{\nu - \case{1}{2} }\left(\frac{m \sigma}{\hbar \tau}  | x| \right) \, ,
\end{align}
which is valid for $\nu > 0$. Here, $K_{\nu}$ denotes the modified Bessel function of order $\nu$~\cite{NIST:DLMF}, which is infinite at the origin but is nevertheless square integrable. The function $\psi (x, \tau)$ is thus continuous, except perhaps at $x=0$. To check the behavior around that point, we use the approximation of $K_{\nu}$ for small values of the argument. This leads
\begin{equation}  
|z|^{\nu - \case{1}{2}} K_{\nu-\case{1}{2}}  \left ( |z | \right ) \approx \frac{\Gamma \left (\nu - \case{1}{2} \right )}{2^{\case{3}{2}-\nu}} + \frac{1}{|z|^{1-2\nu}} \frac{\Gamma  \left (\case{1}{2} - \nu \right )}{2^{\nu - \case{1}{2}}} + O(|z|^{2\nu+1})\, ,
\end{equation}
which shows that the singularity in $\psi (x,\tau)$ thus arises for  $\nu < 1/2$. In summary, when 
\begin{equation}
\label{eq:int}
\case{1}{4} < \nu < \case{1}{2} 
\end{equation}
we get the aforementioned singularity. 

A similar analysis can be performed with the moments of the associated probability density $| \psi(x, t)|^{2}$~\cite{Aiello:2016aa}. The first moment $\langle x \rangle$ is finite and equal to zero when $\nu > 1/2$, whereas the second moment $\langle x^{2} \rangle$ exists provided that $\nu > 3/4$. All this relevant information is concisely summarized in Fig.~\ref{fig1}.
 
\subsection{Quantum trajectories at the singularity}

To explore the physical meaning of the singularity and, more particularly, its dynamical emergence, we resort to the concept of quantum  trajectory.  Apart from providing us with information on the probability density distribution, the wave function $\psi(x,t)$ also contains dynamical information relevant to understand its time evolution. The Bohmian picture   stresses this latter aspect, which manifests as quantum trajectories, which are in compliance with the evolution of the quantum flux~\cite{Sanz:2019aa}.
To this end, one first decomposes $\psi (x,t)$ as $\psi(x,t) = \sqrt{\varrho(x,t)} \exp [iS(x,t)]/\hbar$, which allows us to split up the density information from the phase information encoded in the wave function.
Quantum trajectories are directly related to the local variations undergone by the phase term, $S(x,t)$, according to the so-called Bohmian guiding condition (or local velocity field)~\cite{Holland:1993aa},
\begin{equation}
 \dot{x} = \frac{J(x,t)}{\varrho(x,t)} = \frac{1}{m} \RE \left( \frac{\hat{p} \psi}{\psi} \right)   =  \frac{1}{m} \frac{\partial S(x,t)}{\partial x} ,
 \label{eq3}
\end{equation}
with $\hat{p} = - i \hbar \partial/\partial x$ being the usual momentum operator in the position representation and $J(x,t)$ the probability current density or quantum flux~\cite{Schiff:1968aa}. We stress that Eq.~\eqref{eq3} constitutes a general result that goes beyond any particular interpretation, as it involves quantities that are well defined in any picture of quantum mechanics. 

More importantly, Eq.~\eqref{eq3} explicitly shows the important role played by the phase, not as an indirect effect (e.g., in the appearance of interference features), but as a fundamental quantity that specifies the local dynamics exhibited by the quantum system on each point of the configuration space at each time. This action emerges in the form of the local velocity field that governs the dynamical evolution of the probability density at any time, making it to move from a region to another, to spread out all over the place, or, as it is the case here, to coalesce on a highly localized region at a very precise time.

After all, note that the above local velocity field is what allows us to establish the connection between the probability density, $\varrho(x,t)$, and the quantum flux, $J(x,t)$, according to the well-known transport relation $J(x,t) = v(x,t) \varrho(x,t)$. Quantum  trajectories simply arise after assuming that $v(x,t)$ defines an equation of motion that can be integrated in time, rendering as a result such trajectories. Physically, these trajectories describe the flow of probability at a more local level than the probability density itself does (to some extent, we can say that this latter quantity provides us with a global view of what is going on). A more detailed discussion on the issue can be found in Ref.~\cite{Sanz:2021aa}.

For definiteness, we take the initial state \eqref{eq:psi0}, with $\nu = 1/3$, to ensure a singular wave packet. However, to produce a numerically reliable (and physically more realistic) wave function, instead of the initial ansatz (\ref{eq:psi0}), we consider the following modified one 
\begin{align}
 \psi(x,0) & =  \frac{1}{\sqrt{\mathcal{N}_{\nu}}} \frac{\exp \left (- \frac{i m}{2 \hbar \tau} x^{2}  \right )}{\left (1 + \frac{x^{2}}{\sigma} \right )^{\case{1}{3}}}  \left [ 1 + \tanh  \left ( \frac{x + x_{b}}{\sigma} \right ) \right] 
   \nonumber \\
  & \times \left [ 1 + \tanh  \left ( \frac{x - x_{b}}{\sigma} \right ) \right]  \, ,
 \label{eq1b}
\end{align}
where $x_b>0$. The two smooth step functions represented by the hyperbolic tangents produce a relatively soft decay or cutoff  at  distance $x_b/\sigma$ from the origin, which somehow mimics the effect of a limited aperture with soft boundaries, avoiding the appearance of spurious frequencies associated with a sudden cutoff or Gibbs phenomenon~\cite{Hewitt:1979aa}. Because of the cutoff introduced, it is expected that there will not be time symmetry with respect to $t = \tau$, although the time-evolved of (\ref{eq1b}) will behave close to the exact solution.

\begin{figure*}[!t]
  \centering
  \includegraphics[width=1.95\columnwidth]{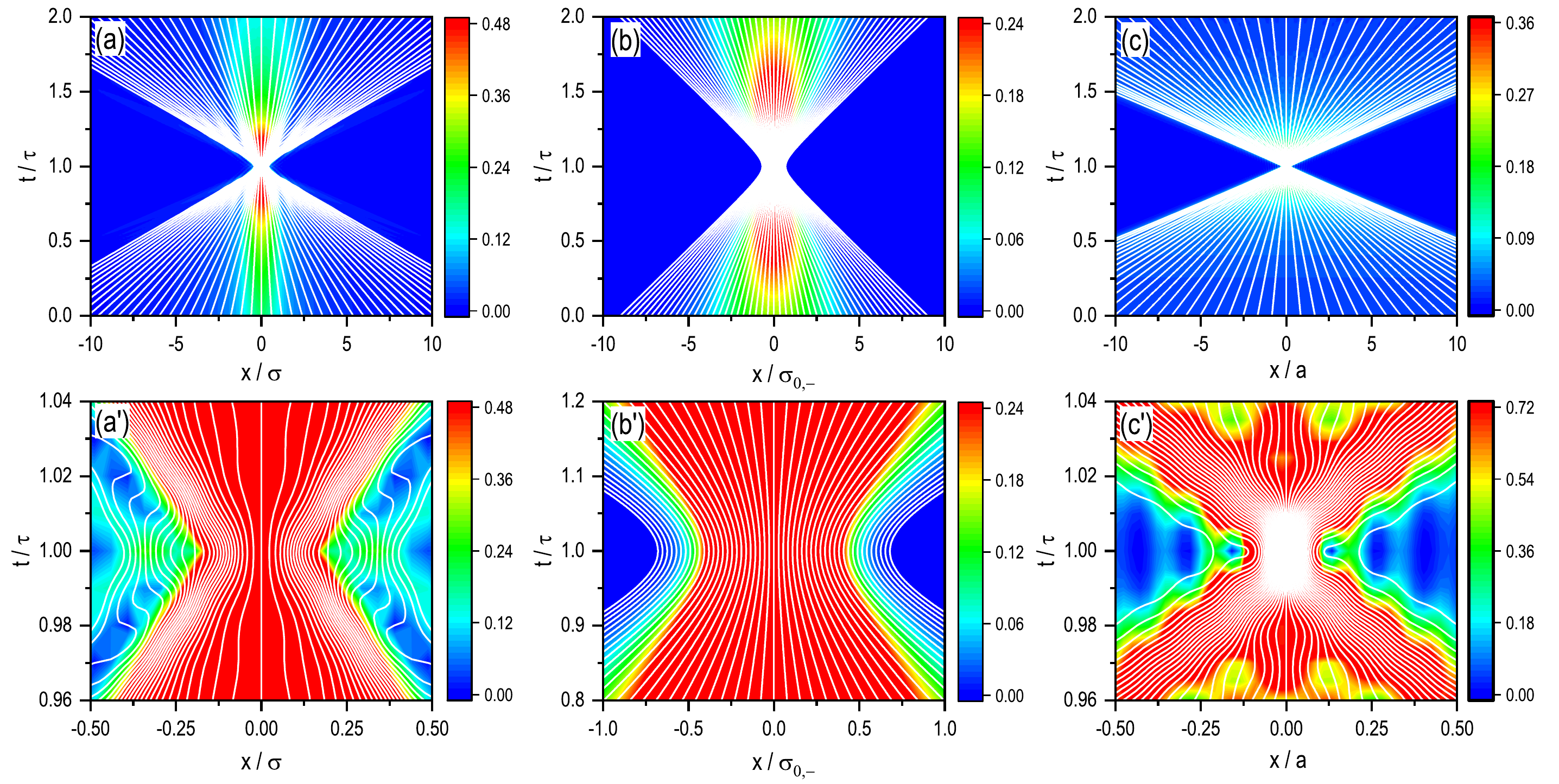}
    \caption{\label{fig2}
 (Top panels)  Quantum trajectories (51)  displayed on top of a density plot describing  the time evolution of the probability associated with  (a) the the wave function \eqref{eq:psi0} with $\nu=1/3$, (b) the Gaussian  \eqref{eq5} with waist width $\sigma_{0,-}$,  and (c) the rectangular wave packet \eqref{eq:rect} with width $a$.  For clarity in the density plot, due to the high values of the probability  density around the singularity, it has been truncated to a tenth of its  maximum value. (Bottom panels) Zoomed version of top panels around the focal region within the time interval where the maximum concentration of probability density is reached.  The whirls in the trajectories denote the appearance and disappearance of nodes  as the wave function approaches its maximum focusing.}
\end{figure*}

We next perform a numerical integration of the evolution \eqref{eq:psit} using a standard pseudospectral method on a spatial mess of size $50 \sigma$ with a total of 1,024 grid points, integrating in time from $t=0$ to $t=2\tau$ with a time step $\delta t = 10^{-3}$, which suffices for our purposes. The numerical solution $\psi(x,t)$ is monitored through both density plots of the corresponding probability density and the associated quantum trajectories. A density plot of the probability density is shown in Fig.~\ref{fig2}a), with a set of 51 trajectories (white solid lines) with equidistant initial conditions between $x/\sigma=-15$ and $x/\sigma=15$ to cover a wide region of the initial probability density. We have chosen $x_b/\sigma = 22.5$. 

As it can be noticed, as time approaches the critical value $\tau$, the swarm of trajectories quickly evolves towards the origin, which turns into a prominent increase of the density within a very narrow spatial region, thus originating the singularity.

This behavior can be better appreciated in the zoomed version around the singular region displayed in Fig.~\ref{fig2}a'). In the same manner, as time proceeds and becomes larger than $\tau$, the swarm of trajectories gets dispersed quickly again. It is worth noting that, while the quantum flux is quite laminar before and after the singularity, as it is indicated by the relative smoothness of the trajectories (they evolve with nearly uniform motion), in the region around the singularity there is a turbulent flow led by the appearance of transient nodes. In their attempt for avoiding these nodes (nodal regions), the trajectories will be forced to undergo a whirling motion.

\section{Singular versus smooth wave packet evolution}
\label{sec:wavepack}

To better understand the singularity, we will next examine a few characteristics of simpler but illustrative cases of smoothly focusing wave packets.

\subsection{Gaussian wave packet}

As it is well known, the evolution of a Gaussian wave packet undergoes an initial boost or acceleration, and then it reaches a stationary linear expansion~\cite{Sanz:2014aa}. Consider the initial normalized Gaussian \emph{ansatz}
\begin{equation}
 \psi(x,0) = \frac{1}{\sqrt{\mathcal{N}_{G}}} \, \exp \left ( - \frac{x^2}{4\sigma_0^2} \right )  \, ,
 \label{eq4}
\end{equation}
where  $\sigma_0>0$ is a real-valued parameter determining the width of the  wave packet and the normalization constant is $\mathcal{N}_{G} = \sqrt{2 \pi \sigma_{0}^{2}}$. Substituting this into the free-space propagator leads to its time-evolved form,
\begin{align}
 \psi(x,t) & = \frac{1}{\sqrt{\mathcal{N}_{G}}}
  \sqrt{\frac{\sigma_0}{\tilde{\sigma}(t)}}  \exp \left [ - \frac{ x^2}{4\sigma_0\tilde{\sigma}(t)} \right ]  \, ,
 \label{eq5}
\end{align}
where the Gaussian complex-valued parameter
\begin{equation}
 \tilde{\sigma}(t) = \sigma_0 \left( 1 + \frac{i \hbar t}{2 m \sigma_0^2} \right) 
 \label{eq6}
\end{equation}
accounts for both the spreading in time of the wave packet, given by
\begin{equation}
 \sigma(t) = |\tilde{\sigma}_t|
  = \sigma_0 \sqrt{1 + \left( \frac{\hbar t}{2 m \sigma_0^2} \right)^2} ,
 \label{eq6b}
\end{equation}
and the development of a space-dependent phase factor.

From the hydrodynamical point of view, the evolution of the above wave function maps onto the trajectories arising from the equation of motion%
\begin{equation}
 \dot{x} =  \frac{\hbar^{2} t}{ (2 m\sigma_0^2  )^{2}}  \,
  \frac{\sigma_0^2}{\sigma(t)^2}\ x .
  \label{eq10b}
\end{equation}
After integration, this equation of motion renders the hyperbolic
trajectories
\begin{equation}
 x(t) = \frac{\sigma(t)}{\sigma_0}\ x(0) .
 \label{eq10}
\end{equation}
From Eq.~\eqref{eq10b}, it is clear that, for $t>0$, the trajectories are ``repelled'' from the region where they are initially confined, namely, the waist of the wave packet, since the sign of $\dot{x}$ directly depends on the sign of $x$ and hence on the corresponding initial conditions. Although the initial expansion is slow, later on, for $t \gg t_s $, with $t_s = 2m \sigma_0^2/\hbar $ being a characteristic spreading time, it becomes essentially linear with time; for $t \sim t_s$, the expansion is accelerated, although at different rates as time proceeds~\cite{Sanz:2012aa}.

All this information is nicely conveyed by the trajectories~(\ref{eq10}), which separate at a rate proportional to their initial distance, $d(0) = |x_2(0) - x_1(0)|$, since $d(t)/d(0) = \sigma(t)/\sigma_0$, where $d(t) = |x_2(t) - x_1(t)|$. Taking into account (\ref{eq6b}), for the same $d(0)$, the largest $\sigma_0$, the slowest the dispersion, and vice versa, in compliance with what is expected in this case.

So far there are no novelties. However, we stress that the above solution is reversible in time, which means that, in the same way that the wave packet undergoes an expansion, it can also be tracked backwards. If the wave packet is then propagated ahead again, it will evolve imploding until reaching a minimum width (waist width), and then expanding again. Taking into account the translational time invariance  of the solutions of the Schr\"odinger equation, if we call $\tau$ the time when waist occurs,  we can define a generalized Gaussian coefficient as $\tilde{\sigma}_{g} (t) = \sigma (t - \tau)$.  In this way the width and the phase of the wave packet at time $t$ are given by   
\begin{equation}
\begin{split}
\label{eq14}
 \sigma_{g} (t) & = \sigma_0 \sqrt{1
  + \left[ \frac{\hbar (t - \tau)}{2 m \sigma_0^2} \right]^2} \, , \\
  \theta_{g} (t) & = \arctan   \left[ \frac{\hbar (t - \tau)}{2 m\sigma_0^{2}} \right] \, . 
\end{split}
\end{equation}
It is clear from these expressions that, at $t = \tau$, we will observe a minimum waist, with $\sigma_{g} (\tau) = \sigma_0$, and zero phase, $\theta_{g} (\tau) = 0$.

\begin{figure}[t]
	\centering
	\includegraphics[width=\columnwidth]{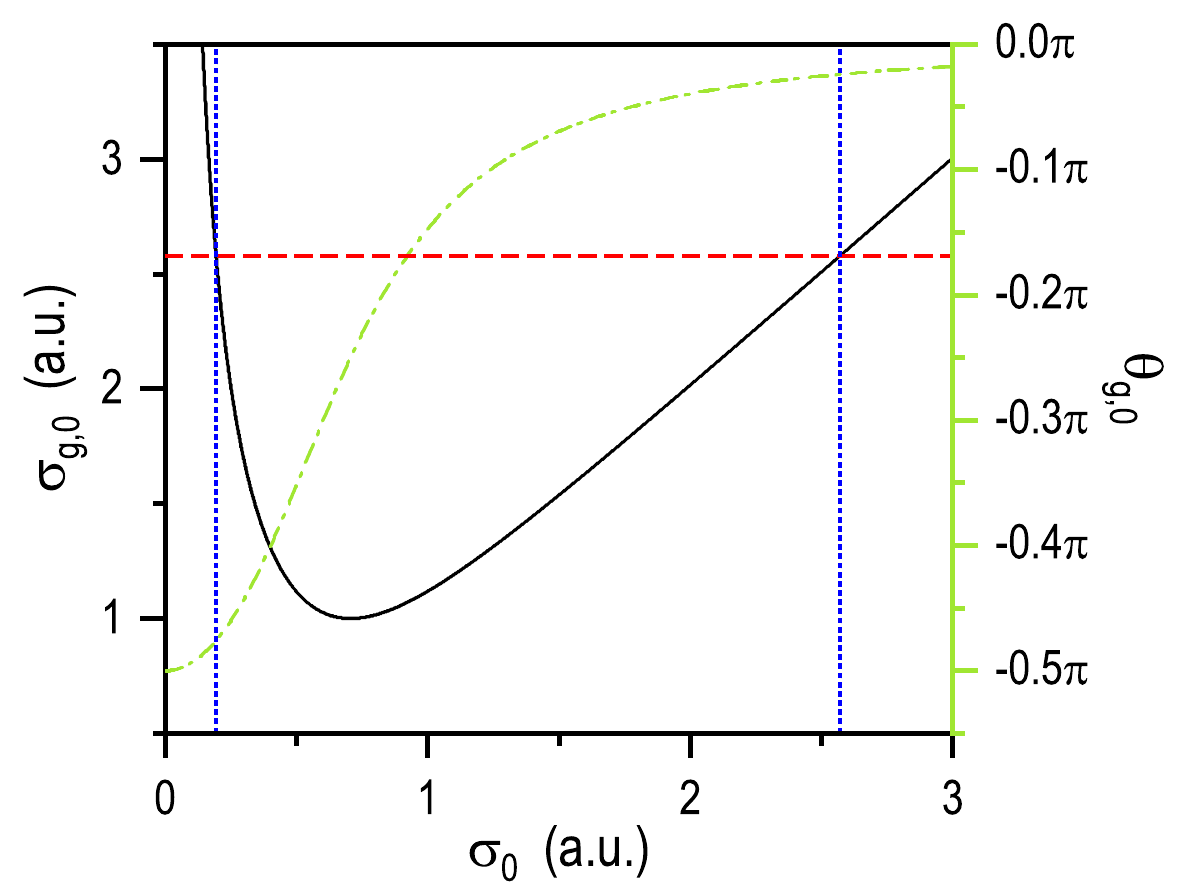}
	\caption{\label{fig3}
 Dependence of the phase (green line) and modulus (black line) of the initial complex-valued Gaussian parameter $\tilde{\sigma}_g$ on the waist width, $\sigma_0$, for $t/\tau = 1$.  The vertical blue dotted lines denote the values of the phase and modulus of  $\tilde{\sigma}_c$ that correspond to Gaussians such that their width at $0.1$ of their  maximum value equals the same value of the probability density corresponding to the wave function.  The horizontal red dashed line   shows that there are always two Gaussian wave  packets with the same initial width, but that lead to two different waist widths (in this case,  $\sigma_g \simeq 2.579$ is associated with $\sigma_{0,+} \simeq 2.571$ and  $\sigma_{0,-} \simeq 0.194$). Despite having the same value for $\sigma_{g}$,  each Gaussian wave packet has a very different initial phase, in particular, $\theta_{g,+} \simeq -0.024\pi$ versus $\theta_{g,-} \simeq -0.477\pi$.}
\end{figure}

Now, contrary to the standard case, we note that there are two factors ruling the expansion dynamics: one associated with the initial width and another one related to a phase, which play opposite roles. If $\sigma_0$ is too large, the phase factor decreases very rapidly, while a small width leads to a prominent phase factor. This dependence is shown in Fig.~\ref{fig3}, where the phase and modulus are separately represented for a better understanding. As it can be seen, $\sigma_{g} (0)$ has a minimum for $\sigma_0 = \sqrt{\tau/2}$, increasing linearly with $\sigma_0$ for large widths and as $1/\sigma_0$  when $\sigma_0$ goes to zero. The associated phase  approaches $-\pi/2$ as $\sigma_0$ decreases, while tends to vanish rapidly as $\sigma_0$ increases above the threshold for minimum $\sigma_{g} (0)$.

\begin{figure}[t]
  \centering
  \includegraphics[width=.95\columnwidth]{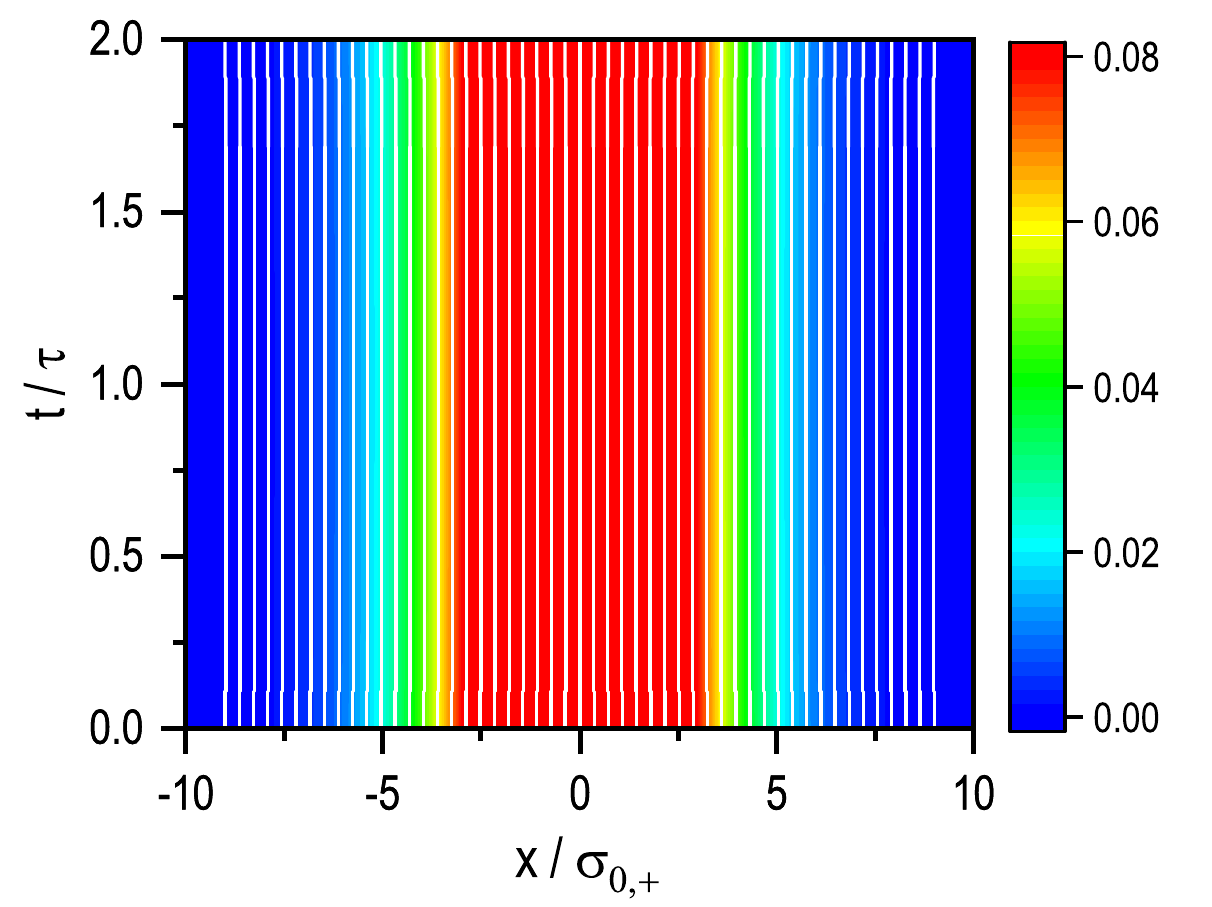}
  \caption{\label{fig4}
 Same trajectories as in Fig.~\ref{fig2}b) for the probability associated with a Gaussian wave packet with waist width $\sigma_{0,+}$.  Note that, because the waist width is relatively large compared to the initial width $\sigma_{g} (0)$, there is no apparent self-implosion (only a very slight narrowing at $\tau$), as it is evidenced by the nearly parallel flux trajectories.}
\end{figure}

From the above discussion, we may now consider the initial Gaussian ansatz as in \eqref{eq4}, but replacing $\sigma_{0}$ with $\sigma_{g} (0)$. The associated time evolution can be directly obtained and leads to the trajectories 
\begin{equation}
 x(t) = \frac{\sigma_{g}(t)}{\sigma_0}\ x(0) .
 \label{eq21}
\end{equation}
As before, these trajectories undergo an initial implosion, until $t = \tau$, and then a subsequent expansion. The question is how important the effect is, particularly taking into account that two different values of $\sigma_0$, as it can readily be noticed from (\ref{eq14}), can be associated with the same initial probability density. These two values  will lead to very different dynamical behaviors. Thus, fixing the value of $\sigma_{g} (0)$, from (\ref{eq14}) we obtain the following two admissible values for the waist width
\begin{equation}
 \sigma_{0,\pm}^{2} =   \frac{1}{2} \sigma_{g}^{2} (0)
  \pm  \sqrt{\sigma_{g}^{4}(0) - \left ( \frac{\hbar \tau}{2 m} \right)^2} \ \, .  
\end{equation}

To quantify the above effect, we consider a Gaussian wave packet with the (initial) width of its probability density at a tenth of the maximum value; i.e., $\varrho_G (s_{\pm},0)/\varrho_G(0,0) = 0.1$, equal to the corresponding value for the (modified) singular wave function (\ref{eq1b}). This yields an initial width for both wave packets given by $\sigma^{2}_{g} (0) = (10\sqrt{10}-1)/(2 \ln 10) \simeq 6.6512$, which gives the waist widths $\sigma_{0,+} \simeq 2.571$ and $\sigma_{0,-} \simeq 0.194$. When compared with the value for $\sigma_{g} (0)$, we notice that while $\sigma_{0,+}$ is practically the same [$\sim 99\%\ \sigma_{g} (0)$], which already indicates a poor dynamics, $\sigma_{0,-}$ is significantly different [$\sim 7.5\%\ \sigma_{g} (0)$] and hence a more relevant dynamical behavior is expected.

\begin{figure}[t]
  \centering
  \includegraphics[width=0.93\columnwidth]{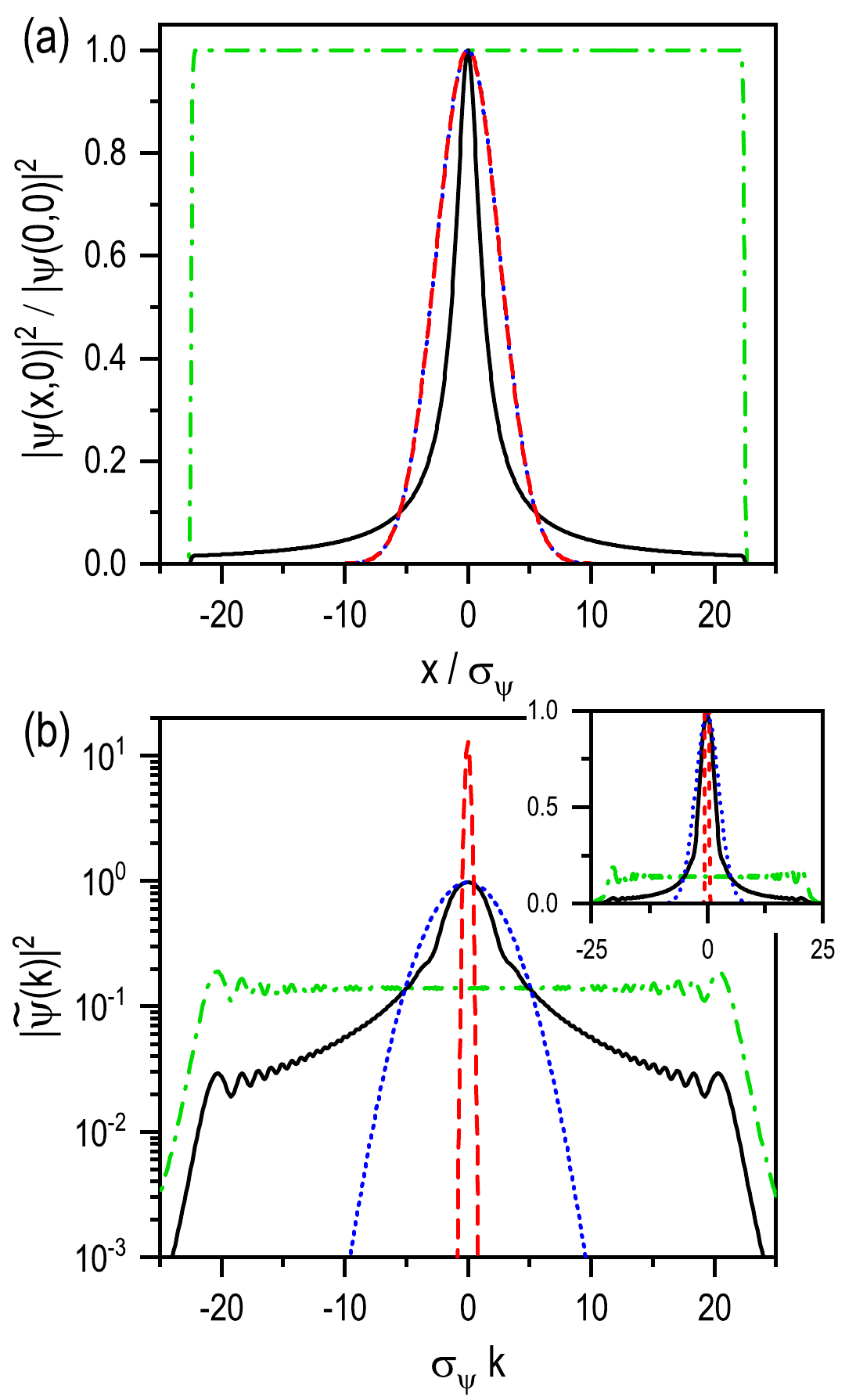}
  \caption{\label{fig5} Probability density in the $x$-position space (upper panel) and in the $k$-momentum space (lower panel) for the singular wave function (black solid line), a Gaussian wave packet with waist width  $\sigma_{0,+}$ (red dashed line), and a Gaussian wave packet with waist width $\sigma_{0,-}$  (blue dotted line), and a rectangular wave packet (green dash-dotted line) for $t = 0$.  The inset shows the same plots on a linear vertical scale. The waist widths for both Gaussians have been adjusted to the width of the probability density for the singular wave function at $0.1$ of its maximum value.}
\end{figure}

The above expectations translate into the results displayed in Fig.~\ref{fig2}b) for $\sigma_{0,-}$.  The characteristic time scale here is $t_{s,-} \simeq 0.15$, about a tenth of $\tau$ and hence with noticeable effects both in the implosion and, afterwards, in the subsequent dispersion. Note here that there is a more important phase contribution, since $\theta_{g,-} (0) \simeq -0.38\pi$, a value closer to the maximum bound for the phase. Nonetheless, unlike the  singular wave packet, here near the singular region the flux is not turbulent,  which is consistent with the fact that the evolution of a Gaussian wave packet is characterized by the absence of nodes.

In  Fig.~\ref{fig4} we plot the reverse case of a Gaussian wave packet for  $\sigma_{0,+}$. We can appreciate that the wave packet remains unaffected, with the flux described by the swarm of 51~Bohmian trajectories being nearly stationary. The characteristic spreading time scale is $t_{s,+} \simeq 26.4 \tau$, which implies that neither the evolution before $\tau$ nor afterwards is going to be importantly affected. Indeed. the initial phase is $\theta_{g,+} (0) \simeq -0.061 \pi$, which already indicates the rather small contribution of the phase factor in the dynamics.

In Fig.~\ref{fig5} we represent the probability densities associated with these initial Gaussian wave packets. Interestingly, these probability densities are indistinguishable in position space, but they are completely different in momentum space: the momentum distribution for $\sigma_{0,+}$ is rather wide, while for $\sigma_{0,-}$ it approaches a Dirac delta function. It is precisely this wider momentum distribution that allows the second wave packet to coalesce toward the origin as the time approaches $\tau$, similarly to the singular wave function, while the first wave packet will remain essentially the same.

\begin{figure}[t]
  \centering
  \includegraphics[width=0.95\columnwidth]{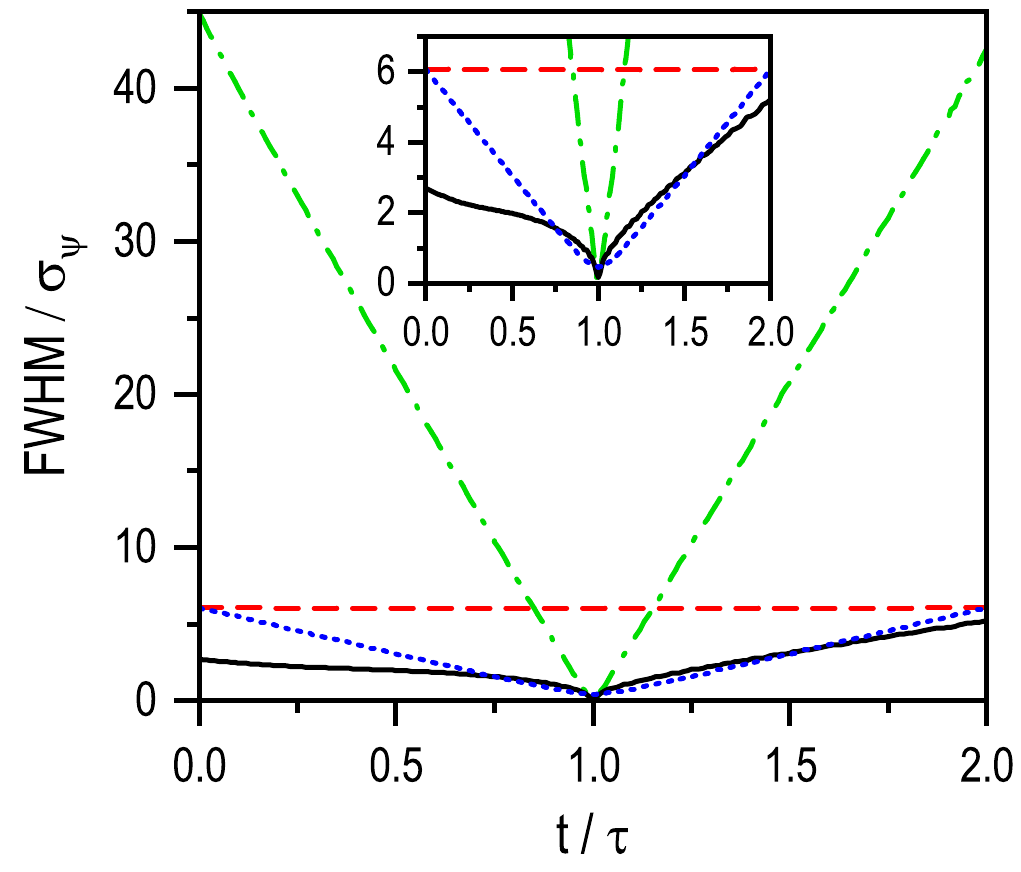}
  \caption{\label{fig6} Evolution temporal of the FWHM for the same wave packets as in Fig.~\ref{fig5}, with the same symbols: the singular wave function (black solid line), a Gaussian wave packet with waist width  $\sigma_{0,+}$ (red dashed line), a Gaussian wave packet with waist width $\sigma_{0,-}$  (blue dotted line), and a rectangular wave packet (green dash-dotted line). }
\end{figure}

\subsection{Rectangular wave packet}

As our last example, we consider a rectangular wave packet~\cite{Mita:2007aa}, with an initial profile
\begin{equation}
\label{eq:rect}
\psi(x,0) = \frac{1}{\sqrt{\mathcal{N}_{r}}} \exp \left (- \frac{i m}{2 \hbar \tau} x^{2}  \right )  \rect_{a} (x) \, ,
\end{equation}
where the rectangle function $\rect_{a} (x)$ is defined as 1 for $|x| \le a/2$ and 0 for $|x| > a/2$ and the normalization constant is $\mathcal{N}_{r} = a$. The time evolution can be found using again \eqref{eq:psit}, finding~\cite{Mita:2007aa}
\begin{align}
\psi(x,t) &  =  \frac{(-1)^{3/4}}{\sqrt{4i \mathcal{N}_{r}}} \exp \left ( \frac{i m}{2 \hbar \tau} x^{2}  \right )  \left \{ \erfi \left [ (-1)^{1/4}  
\sqrt{\frac{m}{2\hbar t}} \left (x - \frac{a}{2} \right )  \right ] \right . \nonumber \\
& - \left . \erfi \left [ (-1)^{1/4} \sqrt{\frac{m}{2\hbar t}} \left (x + \frac{a}{2} \right )  \right ] \right \} \, ,
\end{align}
where $\erfi (x)$ is the imaginary error function and this is valid for $t>0$.  

The wave packet is composed of an infinite number of plane waves. At time $t=0$ these plane waves interfere to give a rectangular shape. As time elapses, the component plane waves travel, both to the right ($k>0$) and to the left ($k<0$), at different phase velocities $\hbar k/2m$. Thus the pattern of the interference of these plane waves gradually changes, resulting in the dispersion of the wave packet.

In Fig.~\ref{fig2}c), we plot the quantum trajectories associated with this evolution. Near the time $\tau$, we appreciate the presence of wiggles for both the singular and the rectangular wave packets, which remind of a nonlaminar flux. Conversely, the  Gaussian profile looks perfectly laminar nearby the singular point.  We recall that a flood in a river occurs because at some point water slows down and the quicker mass of water arriving from behind finds this ``potential barrier" created by the slow water and tries to overcome it. In this case, the wiggles mark somehow a \emph{slower light flow}, so that energy accumulates nearby the singularity and the density grows. 

An alternative way to capture the degree of localization of a wave function
is by studying the behavior exhibited its full width at half maximum (FWHM)~\cite{Garcia-Sanchez:2022aa}. More specifically, this quantity is computed in all cases determining the distance between the two positions, $x_+$ and $x_-$, at which the corresponding probability density reaches half its maximum value at any time; that is
\begin{equation}
 \frac{\varrho(x_{\pm},t)}{\varrho_{\rm max}(x,t)} = \frac{1}{2} .
\end{equation}

Except for Gaussian wave packets, the above equation cannot be solved analytically, so $x_+$ and $x_-$ have been numerically determined on the fly, during the time-evolution of the corresponding wave functions. From this, we obtain ${\rm FWHM} (t) = x_+(t) - x_-(t)$, which is shown in Fig.~\ref{fig6} for the for cases here considered. As it can be noticed, while the FWHM is nearly constant for the Gaussian with waist width $\sigma_{0,+}$, it shows a linear decrease and increase, before and after the waist, respectively, for the Gaussian with $\sigma_{0,-}$. A similar trend is also observed for the square wave function, although the FWHM shows a tiny asymmetry before and after the singularity, which is related to the limitations involved in the numerical method (the spatial size of the grid sets a cutoff for the high spatial frequencies). Finally, for the singular wave function \eqref{eq:psi0}, the  FWHM slowly decrease until $t$ is close to $\tau$, as it can be appreciated in the inset of Fig.~\ref{fig6}. Near this time, the FWHM undergoes a sudden decrease and then increase afterwards; at any later time, the FWHM increase near linearly, in a similar fashion to the Gaussian  with $\sigma_{0,-}$. We notice again a different  behavior between the FWHM dynamics before and after $t = \tau$, which is related to the fact that the wave function considered is not exactly the ansatz~eqref{eq:psi0}), but the truncated version~\eqref{eq1b}. All these characteristics concur with the corresponding probability density and quantum trajectories displayed in Fig.~\ref{fig2}.

\section{Concluding remarks}
\label{sec:conc}

To summarize, we have studied a family of solutions of the Schr\"{o}dinger equation that spontaneously develop a singularity while propagating in free space. Due to the finiteness of these solutions, their singularities do not require a nonphysical infinite amount of energy to manifest. Nevertheless, the local amplitude of the field at a singular point may grow unboundedly. We have given a physical interpretation in terms of quantum trajectories. 

While there is a widespread belief that extreme focusing requires strong nonlinear effects, we have demonstrated that this can be easily achieved with only linear propagation. This promising field enhancement mechanism may foster further interesting research in fields such as electron microscopy or optics.


\begin{acknowledgments}
Financial support is acknowledged to the Spanish Research Agency (Grant No.~ PID2021-127781NB-I00). AA acknowledges support from Deutsche Forschungsgemeinschaft  (Grant No.~429529648-TRR 306).
\end{acknowledgments}


%

\end{document}